\documentclass[fleqn,usenatbib]{mnras}

\usepackage{newtxtext,newtxmath}

\usepackage[T1]{fontenc}
\usepackage{times,ae,aecompl} 

\usepackage{graphicx}	

\newcommand{\be}{\begin{eqnarray}}
\newcommand{\ee}{\end{eqnarray}}

\title[On cosmological bias due to flux magnification]{On cosmological bias due to the magnification of shear and position samples in modern weak lensing analyses}

\author[Duncan, C.A.J., Harnois-D\'{e}raps, J., and Miller, L.]{Christopher A. J. Duncan$^{1}$\thanks{E-mail: christopher.duncan@physics.ox.ac.uk}, 
Joachim Harnois-D\'{e}raps$^{2}$,
Lance Miller$^{1}$
\\
$^{1}$Astrophysics, Department of Physics, University of Oxford, Denys Wilkinson Building, Keble Road, Oxford, OX1 3RH, UK\\
$^{2}$School of Mathematics, Statistics and Physics, Newcastle University, Herschel Building, NE1 7RU, Newcastle-upon-Tyne, UK\\
}

\date{Accepted XXX. Received YYY; in original form ZZZ}

\pubyear{2021}

\begin{document}
\label{firstpage}
\pagerange{\pageref{firstpage}--\pageref{lastpage}}
\maketitle

\begin{abstract}
The magnification of galaxies in modern galaxy surveys induces additional correlations in the cosmic shear, galaxy-galaxy lensing and clustering observables used in modern lensing ``3x2pt'' analyses, due to sample selection. In this paper, we emulate the magnification contribution to all three observables utilising the SLICS simulations suite, and test the sensitivity of the cosmological model, galaxy bias and redshift distribution calibration to un-modelled magnification in a Stage-IV-like survey using Monte-Carlo sampling. We find that magnification cannot be ignored in any single or combined observable, with magnification inducing $>1\sigma$ biases in the $w_0-\sigma_8$ plane, including for cosmic shear and 3x2pt analyses. Significant cosmological biases exist in the 3x2pt and cosmic shear from magnification of the shear sample alone. We show that magnification induces significant biases in the mean of the redshift distribution where a position sample is analysed, which may potentially be used to identify contamination by magnification.
\end{abstract}




\section{Introduction}\label{sec:intro}

As well as inducing distortions in the observed shape of sources, weak lensing also induces a change in the observed size, and consequently flux, of sources. Such changes are highly correlated with the shear and foreground density field, and can therefore induce additional correlations due to selection on flux- or size-dependent measures (such as the measured signal-to-noise of a source) in determining the shear and clustering samples. It is common for the analysis of modern weak lensing surveys to analyse a ``3x2pt'' data vector containing the cosmic shear, galaxy-galaxy lensing and galaxy clustering, each a 2-point correlation in either Fourier/Harmonic space or configuration space. The effect of magnification in the position signal (termed ``magnification bias'') causes additional correlations in the clustering and galaxy-galaxy lensing part of the 3x2pt data vectors, and the impacts of such an effect on cosmological parameter estimation is now well studied \cite[][]{Duncan2014,  Raccanelli16, Cardona16, Lorenz18, Villa18, ThieleDuncanAlonso, Tanidis20, Viljoen21}. These studies show a consistent picture, that indicates that this magnification bias may provide cosmological information on clustering studies, but does not add valuable extra cosmological information when combined with weak lensing shear measurements. Moreover, as it is present in the measured data vector, if it is not included in the fit model then large biases in the inferred cosmological parameters are expected to result. 

In \cite{ThieleDuncanAlonso}, the authors considered individual tomographic data-vector element contributions to the Fisher matrix, to determine which elements gave the largest contribution to the biases in cosmological parameters due to un-modelled magnification. Importantly, it was found that the biggest contributions came from the galaxy-galaxy-lensing signal, and in particular the cross-correlation between the high redshift source sample and all foreground position samples. This section of the galaxy-galaxy lensing data vector is also the highest signal-to-noise measurement in the absence of magnification, and therefore imparts important cosmological information. Moreover, the cosmological bias due to clustering cross-correlations (where magnification is the dominant contribution to the signal) was typically in the opposite direction to those due to the galaxy-galaxy lensing data vector. Together this implies that the magnification of the position sample can not be effectively mitigated by pruning the data vector, and must therefore be accurately modelled. \citet[][]{Duncan2014, Lorenz18} both indicate that the magnification strength must be known to an accuracy of a few percent to avoid biasing cosmological parameters, although in practice the absolute requirement will depend on the intricacies of the analysis not encapsulated in those Fisher matrix based studies, such as photometric uncertainty and form of the likelihood and priors.

The impact of magnification due the selection of the shear sample (termed herein as ``source magnification'') is relatively less well studied, but has been investigated in the galaxy-galaxy lensing sample in \cite{UnruhGGLMag} and in the cosmic shear sample in \cite{Deshpande2020Cosmo, Despande2020Corrs}. In \cite{UnruhGGLMag}, the authors utilise ray-tracing in the Millennium Simulation suite to emulate the magnification of both the shear and position samples in the galaxy-galaxy lensing signal, allowing for either or both forms of magnification to be considered separately in a range of source and lens redshifts. They find that the magnification of the lens galaxies can induce changes to the measured signal of up to $\sim 45\% $ (Fig. 7), and magnification of the source galaxies induces a change of $\sim 3\%$ (Fig 4), with similar sized changes in the inferred mass profiles. 

In \cite{Deshpande2020Cosmo}, the authors forecast the bias caused by magnification of the shear sample in a cosmic shear analysis for \href{https://www.euclid-ec.org}{Euclid}, alongside the effect of reduced shear which induces additional correlations of a similar form to the magnification. They find that individually, the magnification induces typically larger biases in cosmological parameters than the reduced shear alone. The combination of both magnification and reduced shear induce biases comparable to survey precision across all cosmological parameters considered (in a $w_a$CDM cosmology), with the largest bias seen typically in dark-energy parameters (e.g. a bias of $1.3\sigma$ is seen in $\Omega_{\rm DE}$), indicating that this must be modelled as part of a shear analysis with Euclid. They consider the cross-correlation of the magnification and reduced shear with intrinsic alignments, and find that these cross terms are sub-dominant and can be safely ignored, simplifying the modelling requirement. However, they note that since the additional terms in a 2-point shear analysis resulting from the magnification and reduced shear are themselves 3-point correlations, there is additional complexity in the precise modelling of such terms (requiring various assumptions and fitting formula), and additional computational complexity which can impact the ability to run long cosmological MCMC chains \citep[][]{Deshpande2020Cosmo, Despande2020Corrs}. In \citet{Deshpande2020Cosmo} it is therefore recommended that forward-modelling approaches utilising likelihood-free inference such as DELFI \citep{DELFI1, DELFI2, DELFI3} should be considered, to mitigate the impact of modelling magnification and reduced shear.

Historically, analyses of the data from current-generation weak lensing surveys have been safely able to ignore the magnification effect. However, with the increasing statistical precision with which cosmological parameters can be inferred as more data is taken, the impact of magnification of the position sample is now routinely taken into account when analysing the galaxy-galaxy lensing \citep[e.g.][]{DESY3GGL} and combined 2x2pt \citep[e.g.][]{DESY32x2pt} and 3x2pt analyses \citep[e.g.][]{KIDS1000Cosmology}. As shown by the literature detailed above, next generation surveys such as \href{https://www.euclid-ec.org}{Euclid} and \href{https://www.lsst.org/category/weak-lensing}{LSST} must also routinely model the magnification as part of their analyses. Moreover, due to the increased sensitivity of these future surveys to magnification, the analyses of these data-sets may need to go beyond the current modelling where the magnification is assumed well-known, and potentially allow for simultaneous inference of the magnification alongside cosmological parameters at the expense of cosmological precision \citep[][show that marginalisation of magnification strength for magnification of the position sample induces O(10\%) reduction in parameter constraints, reducing to O(1\%) on application of a $10\%$ prior.]{Lepori}

Each of the studies above give vital information on the impact of the magnification on each individual component of the weak lensing data vector, however to date there has been no investigation which considers the impact of selection effects on both the shear and position samples in a magnification field in combined clustering, galaxy-galaxy lensing (2x2pt) and shear analysis (3x2pt). In this paper, we consider the impact of both magnification of the shear sample and position sample in a photometric 2x2pt and 3x2pt weak lensing analysis through emulation of the signal in the SLICS \citep{SLICS_1} simulations. We consider the impact on cosmological parameters, galaxy bias and systematic uncertainty in the photometric redshift distributions, modelled as shifts in the mean of the distributions for each tomographic bin. Section \ref{sec:theory} details the origin of the extra correlations due to sample selection in a magnification field; Section \ref{sec:sims} describes how these correlations are emulated with SLICS and Section \ref{sec:results} presents the results of a model cosmological inference.

\section{Additional correlations due to magnification}\label{sec:theory}

Gravitational lensing by a distribution of mass between a luminous source and the observer changes the source's observed position, size and brightness, as well as its shape. The former three effects are commonly grouped under the term "magnification", labelled by a magnification factor field $\mu$ related to the local weak lensing convergence $\kappa$. The impact of a local magnification on a population of sources is commonly summarised as 
\begin{eqnarray}
    n(<S)   &=& \mu^{\alpha-1}n_0(<S) \nonumber\\
            &\approx& (1+2[\alpha-1]\kappa)n_0\nonumber\\
            &=& (1+q\kappa)n_0 \label{eqn:mag_number_density}
\end{eqnarray}
where $n$ and $n_0$ represent the lensed and intrinsic number density of sources respectively, and $q=2(\alpha-1)$ denotes the relative change in number density due to magnification bias. In the second step, we have linearised the magnification pre-factor (i.e. applied the ``weak lensing'' limit). Where a hard cut in flux is applied, the magnification strength $\alpha$ may be be related to the distribution of sources, averaged across the sky, at the faint limit of the survey as
\begin{equation}
    \alpha = 2.5\left.\frac{{\rm d} \log_{10} n(>m)}{{\rm d}m}\right|_{m=m_{\rm faint}},
\end{equation}
with apparent magnitude $m$. Similar relations apply for a hard cut at bright magnitudes, and where hard selections are applied on measurables that are altered by a magnification field (such as source size, where a selection can be imposed through star-galaxy separation for example). However, these extra contributions can often be combined into an effective magnification strength such that the formalism presented above is still valid to first order. Note however that in doing so, the strength at the faint end and at the bright end are also weighted according to the number density at each end, and therefore for widely separated bright and faint cuts the bright cut is typically a small perturbation on the total strength which is dominated by the faint limit.
In practice, modern surveys will impose a selection of sources which correlates with magnitude (e.g. through signal-to-noise considerations) which is therefore not a hard cut at either faint or bright magnitudes. \cite{HildebrandtMagSystematics} showed that provided the measurement of the magnification strength is done on a sample which is complete at these limits (e.g. from a deeper observation such as Euclid-deep) then this approximation is valid.

Modern weak lensing surveys will consider not only measurements of galaxy shape, but also measurements of local galaxy position, so that three observables
\begin{eqnarray}
    w^{ij}(|\theta|) &=& \langle\delta^i(\theta)\delta^j(\theta+|\theta|)\rangle\;\;\;{\rm (Clustering)}\\
    \xi^{ij}(|\theta|) &=& \langle\gamma^i(\theta)\gamma^j(\theta+|\theta|)\rangle\;\;\;{\rm (Shear)}\\
    \gamma_t^{ij}(|\theta|) &=& \langle\delta^i(\theta)\gamma^j(\theta+|\theta|)\rangle\;\;\;{\rm (GGL)}\label{eqn:base_ggl}
\end{eqnarray}
can be measured and compared to theory. Here $\delta$ labels the local number over-density at sky position $\theta$, and $\gamma$ the shear, and roman indices $i,j$ label redshift bins. The combination of these observables can provide more precise and accurate cosmological constraints through the addition of extra cosmology-dependent information, as well as self-calibration of nuisance parameters which may be degenerate with cosmology such as the galaxy bias \cite[see e.g.][]{Joachimi10}.

In the presence of a magnification field, the samples of both the shear and position are affected locally, leading to additional terms in the measured correlations. The magnification of the position sample leads to the following additional contributions \citep[e.g.][]{Duncan2014}
\begin{eqnarray}
    \Delta w^{ij} &=&  q^{j}\langle \delta^{i}_g \kappa^{j} \rangle + q^{i}\langle\kappa^{i}\delta^{j}_g \rangle + q^iq^j\langle\kappa^i\kappa^j\rangle\label{eqn:pos_mag_corrs_cl}\\
    \Delta \gamma_t^{ij} &=& q^i\langle\kappa^i\gamma^j\rangle\label{eqn:pos_mag_corrs_ggl}
\end{eqnarray}
where $\kappa$ is the local convergence field, and $\delta_g$ labels the intrinsic number over-density fluctuation in the absence of magnification. 

Each additional term has a physical interpretation which can aid in understanding their origin. In Eqn \ref{eqn:pos_mag_corrs_cl}, the first (and second) terms correspond to correlation between the clustering of the foreground and the magnification of the background (and vice-versa), which is non-zero only due to the local matter environment of the foreground bin. Note that the second term is zero where $i<j$, except where there is overlap in the redshift distributions (as matter in the background of a source cannot lens it). The final term is a pure magnification term, and labels the correlation due to magnification of both the foreground and background, due to shared large-scale structure (LSS) in between the foreground bin and the observer. 

Eqn \ref{eqn:pos_mag_corrs_ggl} gives the correlations between the magnification of bin $i$ and the shear of bin $j$, due to shared LSS in between the observer and the foreground bin. Note that this can be non-zero where $j<i$ even without redshift bin overlap, where Eqn \ref{eqn:base_ggl} would have zero signal, and as such these redshift bin combinations can consist of a galaxy-galaxy measurement which is dominated by magnification. This may be particularly problematic if those correlations are used to put constraints on an intrinsic alignment model.

The magnification of the shear sample leads to the further addition of \citep[see e.g.][]{Schmidt09}
\begin{eqnarray}
    \Delta \xi^{ij} &=& \langle [\delta_g^i+ q_\gamma^i\kappa^i]\gamma^i\gamma^j\rangle + \langle\gamma^i[\delta_g^j + q_\gamma^j\kappa^j]\gamma^j\rangle\label{eqn:src_mag_gg}\\
    \Delta \gamma_t^{ij} &=& \langle [\delta^{i}_g +
    q^i\kappa^i]\delta^j_g\gamma^j\rangle + 
    q_\gamma^j\langle[\delta^i_g + q^i\kappa^i]\kappa^{j}\gamma^j\rangle,
\end{eqnarray}
resulting from an effective change in the measured localised shear due the local change in number density of sources as
\begin{equation}
    \gamma\to \gamma(1+\delta_g + q_\gamma\kappa).
\end{equation}  
We distinguish the localised re-weighting of the shear sample $q_\gamma$ from that of the position sample for clarity, but note that the effect of magnification in both terms is given by Eqn \ref{eqn:mag_number_density}.
To interpret the physical origin of these extra terms, it is important to understand that they are concerned with correlations between foreground and/or background shear samples whose number density is modified by magnification due to foreground LSS as well as by their local dark matter environment. Thus, we see that for the cosmic shear signal (Eqn \ref{eqn:src_mag_gg}) the first term gives the correlation between the shear of the background sample and  the number-boosted foreground shear sample, with the term in brackets giving a weighted number density including local intrinsic fluctuations and magnification by LSS. The second term similarly correlates a foreground with a number-boosted background. Notably then, the additional terms preserve symmetry in the redshift bin labels. Of these terms, correlations between terms of the form $\langle \delta_g \gamma\gamma \rangle$ denote a form of source-lens clustering requiring a significant overlap between the lensing kernels and the redshift distribution, and therefore expected to be small for sufficiently narrow redshift bins \citep[see the discussion in][for further detail]{Schmidt09}. The remaining terms of the form $\langle\kappa\gamma\gamma\rangle$ correlate the shear and convergence fields sourced from the same foreground LSS.
These latter terms therefore identify the addition correlation that comes from the fact that we select relatively more sources around areas with positive magnification (for $\alpha > 1$ ), and that these also correspond to more highly (tangentially) sheared sources. Note the importance of the size of the magnification strength: for $0 < \alpha < 1$,  we would under-represent highly sheared sources around regions with positive magnification.

Similarly, for the additional galaxy-galaxy lensing terms, the first term correlates the number-boosted foreground with the shear of the background weighted by the number density due to intrinsic clustering, and therefore encompasses source-lens clustering and is expected to be subdominant. The second correlates to the background shear weighted by the magnification-boosted background, thus correlating the local matter density at the lens and the background convergence and shear field, as well as the convergence of the lens with the lensing of the source due to the same foreground LSS. The interpretation of this is similar to that of the cosmic shear signal, where we gain additional correlation as the foreground lensing mass causes over- or under-represention of sources in a way which is correlated to their tangential shear.

In practice the source shear is not directly inferred in weak lensing analyses, but rather the reduced shear $g = \gamma/(1-\kappa)$ is measured. As noted in \cite{Schmidt09, Deshpande2020Cosmo}, the reduced shear correction induces additional terms in the 2-point functions which have the same form as the magnification, provided we work in the weak lensing limit so that the reduced shear correction can be linearised in conjunction with the magnification. In this case, the effective shear re-weighting can be altered to 
\begin{equation}
    q_{\gamma} \to 2(\alpha-1) + 1,\label{eqn:reduced_shear_weight}
\end{equation}
where the first term is the magnification and the second the reduced shear. 

We see immediately that the position correlations are sensitive only to magnification in the position sample, shear correlations sensitive only to the shear sample, and the galaxy-galaxy-lensing is sensitive to magnification of both the shear and position sample. Further, we emphasise that the magnification of the shear sample induces additional correlations which are 3-point / bi-spectrum terms, and as a result are expected to be an order of magnitude smaller in the weak lensing limit than those induced by the magnification of the position sample, which are 2-point / power spectra. Therefore, it is expected that cosmological parameters are more sensitive to the magnification of the position sample than the shear sample. However, the computational complexities in modelling the bi-spectrum terms \citep[see e.g.][for a description]{Deshpande2020Cosmo, Despande2020Corrs} means that these terms can be expected to be more difficult to model for inclusion in the model vector which is compared to data measurements as part of cosmological fits.

\section{Emulating the magnification with weak lensing simulations}\label{sec:sims}

We capture the effect of magnification by measuring the $3\times$2-pt statistics in mock galaxy catalogues, in which the magnification terms can be turned on and off at will.  The signal sought receives a strong contribution from small, highly non-linear scales, while at the same time we need to have access to the large, linear scales, in order to avoid over-estimating the impact of magnification on a realistic cosmic shear measurement. We meet these combined requirements by exploiting the `high-resolution' Scinet LIght-Cone Simulations \citep[SLICS-HR, introduced in][]{SLICS_1}, which resolves $k$-modes well beyond 10 $h{\rm Mpc}^{-1}$. 

The SLICS-HR consists of 5 fully independent $N$-body runs carried out by {\sc cubep$^3$m} \citep{cubep3m}, in which  1536$^3$ particles were evolved from $z=99$  to $z = 0$ in a $505 h^{-1}{\rm Mpc}$ box. Similar to the methods detailed in \citet{SLICS}, mass sheets and dark matter haloes were extracted at run time and subsequently post-processed into 100 deg$^2$ light-cones. Specifically, we construct 25 {\it pseudo}-independent light-cones by randomly shifting the origins of the mass sheets, by shuffling the axis along which the density fields are collapsed into mass sheets, and by selecting at random which of the 5 independent runs contribute to a given redshift segment of the light-cone; as for the main SLICS suite, there are 18 redshift segments per light-cone, each $(505/2)\, h^{-1}$Mpc thick. The mass sheets are then ray-traced such as to compute convergence and shear maps at 18 redshifts in the range $z \in [0.0 - 3.0]$, from which lensing quantities can be interpolated. 

Dark matter haloes are populated with an Halo Occupation Distribution (HOD) model following the LSST-like method described in \citet{SLICS}. To summarise, the model is based on a conditional luminosity function, in which the mean number of satellite and central galaxies depends on the mass of the host halo and on the luminosity range, which is set to $[26.7<M_r<18.0]$.

All HOD parameters are made luminosity dependent and vary as a function of redshift, halo mass and a threshold $L_{\rm min}$. We refer the reader to section 3.7 of \citet{SLICS} for a full description.
In the end, every galaxy is assigned a position, redshift, shear and convergence, resulting in a  number density of $n_{\rm gal} = 25.8$ gal arcmin$^{-2}$. These galaxies are subsequently split in tomographic bins and used either as weak lensing sources or foreground lenses, in order to study the $3\times2$-pt data vector.

\begin{figure}
\includegraphics[width=0.45\textwidth]{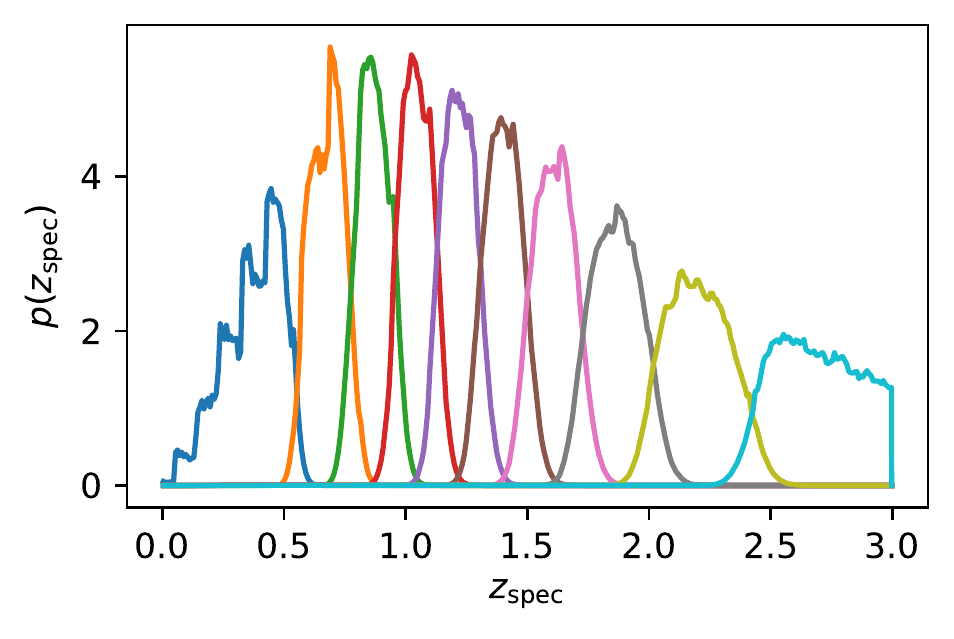}
\caption{Tomographic redshift distributions from the SLICS after convolution with a Gaussian photometric uncertainty. Further details on spectroscopic redshift assignment can be found in \citet{SLICS}. }\label{fig:nz}
\end{figure}

Photometric redshifts are assigned to each source galaxy in the simulation catalogue by sampling from $\mathcal{N}(\mu=z,\sigma = 0.02[1+z])$, where $z$ is the true redshift from the input catalogue. The simulations used therefore model basic typical photometric redshift uncertainty but do not model outliers or non-Gaussianity. The simulated catalogues as analysed are therefore not constructed to mimic closely the properties of any given future weak lensing survey, but instead to provide a simulacrum of a typical Stage-IV survey. The simulated catalogues are binned into 10 tomographic redshift bins, with an equal number of sources in each bin. Fig. \ref{fig:nz} shows the resulting redshift distributions.

The SLICS simulations do not provide magnified source magnitudes and positions, and therefore the magnified correlation functions cannot be directly measured from the simulations. However, a redshift dependent convergence map is supplied with the simulations, allowed for the allocation of a source specific convergence. From this, we emulate the magnification contribution, by re-weighting each source according to Eq. \ref{eqn:mag_number_density} (or using Eqn \ref{eqn:reduced_shear_weight} when considering the reduced shear sample) when measuring each correlation function with {\sc TreeCorr} \citep{treecorr}. Correlations are measured on the simulated data using logarithmic binning on 8 bins between $\theta = (1',100')$.

The magnification strength is estimated utilising the fitting for the magnitude distribution in \cite{KannawadiAlphaFitting} (Eqn 16), evaluated as a local power law at a limiting magnitude of $m=24$, giving $\alpha = 1.3$ for the baseline application. This is applied to all redshift bins for both the shear and position samples, but we note that in practice the inferred value of alpha is strongly dependent on the survey selection, and expected to vary with redshift. This analysis therefore provides parameter bias significances which are indicative of those expected in a typical Stage-IV survey, but does not provide accurate biases for any specific survey. We note that in \citet{Lepori}, the authors inferred a redshift-dependent magnification strength from \emph{Euclid} Flagship simulations which varied from $\alpha\approx 0.06$ at low redshift to $\alpha\approx 2.5$ at high redshift, and the value of $\alpha = 1.3$ chosen here corresponds roughly to their inferred value at $z\approx 1.15$. Similarly \citet{Deshpande2020Cosmo} utilise a magnification strength inferred from a fit to the SDSS $r-$band ranging from $\alpha\approx 0.5$ to $\alpha\approx 2.5$ at low and high redshift respectively, and our choice corresponds to their inferred value at $z\approx 1$. The baseline absolute magnification bias correlations emulated in this work are therefore typically smaller than those in \citet{Lepori, Deshpande2020Cosmo} at low and high redshift, and larger at intermediate redshifts. The reduced shear reinforces the magnification for any $\alpha > 1$, and thus in \citet{Deshpande2020Cosmo} the magnification and reduced shear partially cancel for $z\lesssim 0.8$. Consequently, where reduced shear is included, the additional correlations emulated here are typically larger at low redshift than in \cite{Deshpande2020Cosmo}. We therefore also consider as a special case the propagation of the redshift-dependent $\alpha$ values of \citet{Lepori} (L21). In this case, the L21 $\alpha$ values are interpolated by mean redshift of the bin. Where the bin is at higher redshift than their maximum we do not extrapolate, but assume the value of $\alpha$ is constant beyond this range since their derived values are approximately flat in the three highest redshift bins.  

The re-weighting mimics a change in the representation of each source in the sample according to its convergence field: a source with a large positive convergence will effectively have its numbers boosted due to magnification (since $\alpha > 1$), whilst one with negative convergence will have its number effectively depleted. This re-weighting can be applied to either or both of the position and shear catalogues, giving flexibility to activate individual elements of the magnification contribution to the data vector to identify the importance of the magnification of each sample individually or together. We impose the linearity assumption in the re-weighting, and therefore cannot test the sensitivity to this assumption, but note that this should be explicitly tested in future work.

\begin{figure*}
\includegraphics[trim = 0  105 0 22, clip, width=0.92\textwidth]{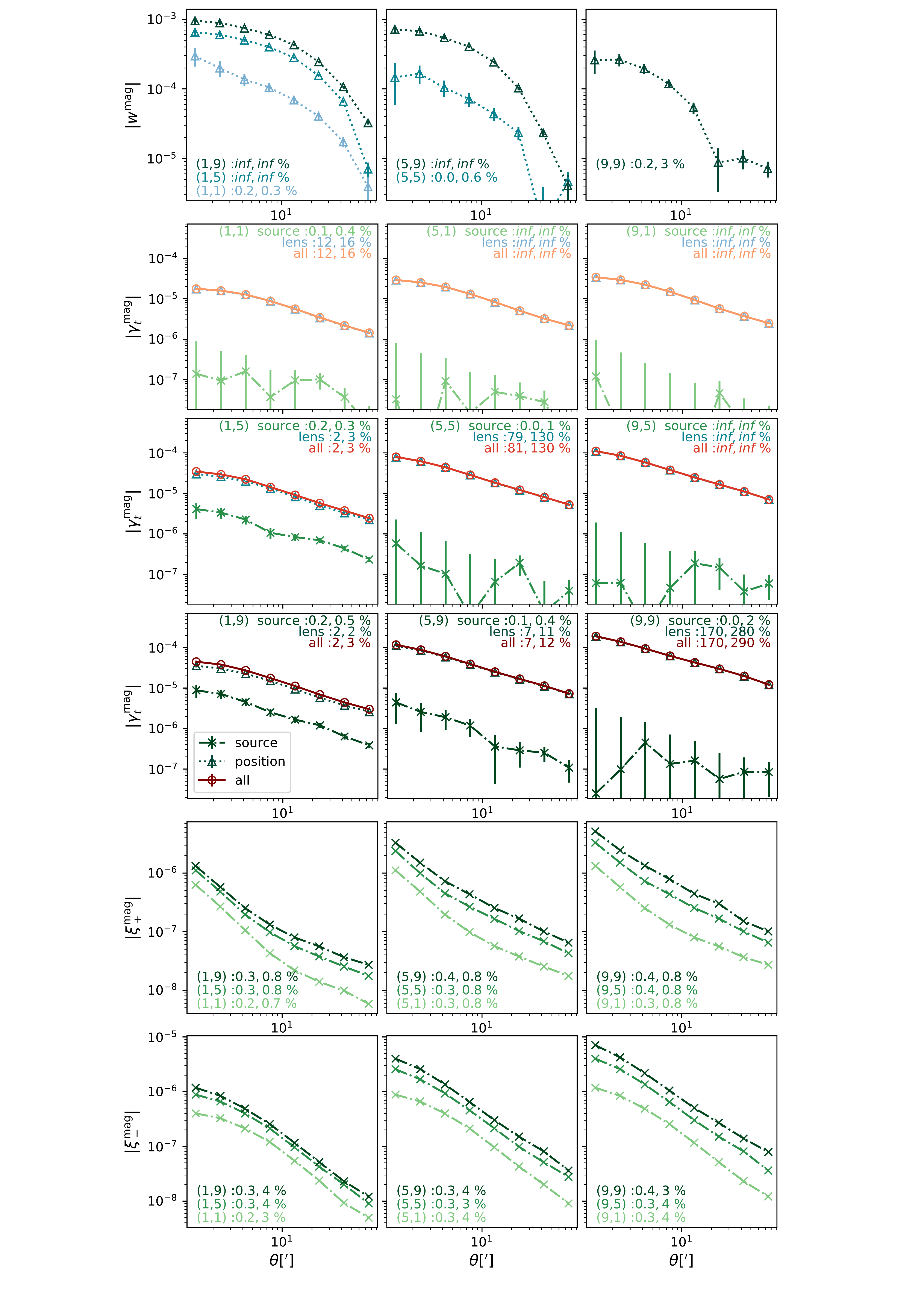}
\caption{Magnification contributions to the data vector as emulated and measured in the simulations (not including reduced shear). Three redshift bins are shown for clarity. Foreground redshift increases left-to-right, and background redshift increases according to darker colour-scale. Position magnification is given by blue, dashed and triangles. Source magnification by green, dot-dashed and crosses. All contributions are given by red, circles and solid. Insert text gives minimum and maximum percent contribution of magnification to un-magnified signal (e.g. $w^{\rm mag}/w$) and numbers in brackets label foreground and background redshift bins.``inf'' values indicate total signal is comprised of only magnification correlations. Errors are the standard error on the mean across light cones.}\label{fig:sims_magnification}
\end{figure*}

Figure \ref{fig:sims_magnification} shows contributions to each element of the data vector according to magnification of either sample (labelled ``position'' for magnification of the position sample, and ``source'' for magnification of the shear sample) or both (labelled ``all''). Each plotted correlation is the mean of the signal measured across each of the simulated light-cones, and the plotted error is the standard error on the mean across all light-cones. Note that for clarity we show combinations of only 3 redshift bins, however all 10 bins are measured and used in the subsequent fits. Inset text gives the minimum and maximum relative fraction of the magnification terms to the un-magnified correlations as a percent value. The clustering correlations assume symmetry in redshift bin labels. The galaxy-galaxy-lensing correlations are not assumed symmetric in redshift bin, and the upper right corner of that section of the plot details the case where the mean of the foreground sample (position) is behind the background sample (shear). These panels are therefore dominated by magnification\footnote{Particularly where the position sample is far enough behind that there is no overlap in the redshift distributions between bins, such as the 9,1 bin combination, and note that outlier fractions are not modelled in the simulations}, and are included here for visualisation purposes, although only bin combinations with the mean redshift of the shear greater than or equal to that of the position is included in the subsequent fits (this is discussed further in Section \ref{sec:inference}). 

The clustering correlations (top row) only contain a contribution from the magnification of the position sample by definition. We see that the amplitude of the magnification bias in these terms increases with increasing redshift of the background sample, and increases with increasing difference in redshift between the foreground and background. Moreover, as the separation between foreground and background increases, the intrinsic (un-magnified) correlation decreases, so that the relative fraction of the magnification signal increases rapidly. Note however that outliers are not modelled here, and that outliers could cause the intrinsic correlation between redshift-separated position samples to be non-zero. The relative size of the magnification terms in the auto-correlations increases with the redshift of both the foreground and background samples, due to an increase in the size of the magnification contribution (in this case, the magnification of both positions samples by common foreground LSS increases) and the reduction in amplitude of the intrinsic clustering signal. 

The shear correlations (bottom two rows) contain only contributions due to the magnification of the shear sample. We see that the amplitude of the correlations due to magnification increases both with increasing redshift in the foreground and background samples. Across the angular separations considered, the relative contribution of magnification to the $\xi_+$ signal (typically sub-percent) is smaller than that of $\xi_-$ signal (typically a few percent).

The galaxy-galaxy lensing correlations contain contributions due to both the magnification of the position (lens) sample and the shear (source) sample. For all measures and all scales, we see that the correlations due to the magnification of the shear sample is sub-dominant to that of the position sample, and its relative contribution is largest where the position sample is at low redshift. In all cases where the shear sample is at higher redshift than the position sample, the source magnification gives a sub-percent effect on the measured correlation. The magnification of the position sample can be a dominant effect in the measured correlation function, with magnification contributions exceeding the intrinsic correlations along the diagonal at high redshift. The relative size of the magnification correlations rises with an increase in the redshift of both samples, and as the redshift separation between the lens and source decreases. This is due to two effects: the intrinsic correlations decrease due to the decrease in the lensing efficiency between close lens halos and source sample; and the increase in the magnification signal due to magnification of both the lens and source samples by the same foreground LSS. This suggests that analyses using the galaxy-galaxy lensing signal at high redshift will be more sensitive to the magnification, as found in \cite{ThieleDuncanAlonso}.

The correlations measured in this way from the SLICS simulations do not contain a contribution from intrinsic alignments (IA), and therefore the analysis presented here does not contain any cross-correlation between the IA and the magnification terms. We note that in \cite{Deshpande2020Cosmo} (Section 4.4) it was found that the cross correlation of the magnification of the shear sample and the IA was sub-dominant to the other terms in the shear-shear correlation in the bias of cosmological parameters. 

\subsection{Parameter Inference}\label{sec:inference}

We model the likelihood as a Gaussian, whose log-posterior is given by the following form
\begin{equation}
    \mathcal{P}(\theta|\mathcal{D}) = -0.5(\mathcal{D}-M(\theta))^TC^{-1}(\mathcal{D}-M(\theta)) + \log(\pi(\theta)),
\end{equation}
up to a parameter-independent additional factor\footnote{Assuming no parameter dependence in the covariance.}, where $\mathcal{D}$ is a vector of all data points to be fit and $M$ the vector of model values, both including the correlation functions across observable, redshift bin and scale, for the set of model parameters $\theta$ including cosmological and nuisance parameters, $C$ is the covariance and $\pi$ is the prior on model parameters, and the evidence is ignored.

The covariance matrix in constructed from an analytical calculation presented in \citet{cosmoSLICS} and we refer the interested reader to \citet{KIDS1000Methodology} for more details on the general calculations.  Constructed specifically for this work, our covariance matrix contains the Gaussian, the connected non-Gaussian and the super-sample covariance terms for a 15,000 deg$^2$ survey, and assumes the same cosmology and $n(z)$ as the simulations. The input matter power spectrum is computed from {\sc Halofit} \citep{Halofit2012}, and the shape noise (per component) and galaxy density are given by $\sigma_{\epsilon}=0.28$ and $n_{\rm gal}=26.0  \,\, {\rm gal \, \, arcmin}^{-2}$, respectively. The use of source-specific weights induces an extra effective selection which may be redshift, magnitude and density dependent (discussed in Section \ref{sec:limitations}) and alter the effective number density. The value for $n_{\rm gal}$ chosen here corresponds to the effective number density in an LSST-like survey \citep{2013MNRAS.434.2121C}.

We next fit the theory to the simulations in two steps. In the first step, we fit for non-linear bias terms directly to the set of correlations as measured from the simulations, where magnification has not been emulated. This step is necessary since a bias model is required in the covariance matrix, and simulated galaxy catalogues are constructed from an HOD model with a scale and redshift dependence that is unspecified {\it a priori}. This step is discussed in further detail in Appendix \ref{sec:app_fit_nl_bias}, but note that in that case the covariance is evaluated for a different angular survey window size to that given above. Theory curves are constructed using the {\sc CCL} library \citep{CCL} utilising CLASS \citep{CLASS}.

In the second step, the data vector is constructed as the sum of the noise-free (un-magnified) theory fit in fitting step 1, and the simulation-derived magnification terms. The simulation-derived magnification terms are constructed as the difference between the measured correlation functions with emulated magnification and those without magnification emulated. Whilst each set of correlation functions are noisy due to the finite size of the simulations, a large amount of noise cancellation occurs in the difference, and therefore the data vector is approximately noise-free. As a result, estimated parameters are approximately noise-free, and thus  can be used to derive accurate biases on cosmological parameters conditioned on which magnification terms are included in the data vector and the form of the fit model, whilst parameter uncertainties are dictated primarily by the form of the covariance.

As the number of data elements is significantly larger than in previous generations of this type of calculation (i.e. the total number is 880 data points for the full 3x2pt analysis after scale and redshift-bin cuts), we ran into an unforeseen numerical instability in which the full covariance was no longer positive definite. This prevented us from adequately inverting the matrix as per our requirement in the likelihood evaluation. As we could not find the source of this problem, we carried out a block-by-block inspection and found that removal of the galaxy-galaxy lensing correlations where the position sample is at higher redshift than the shear sample improved this for the first fitting step. These correlations are therefore removed from all fits. Whilst correlating foreground shear and background position may be used to infer an intrinsic alignment (IA) model, we note that in \cite{ThieleDuncanAlonso} it was found that the contribution to the bias due to magnification in these bins was subdominant to the case with a background shear sample, and since we do not model the IA effect we could not anyway interpret the impact on the inferred IA model due to the magnification in these bins. We also found that the SSC term used in the second fitting step imparts non-positive definite component in the full covariance; as this term is sub-dominant for our survey configuration, we decided to simply remove the offending SSC contributions. Note that the SSC contribution is included in the fits in step 1 (Appendix \ref{sec:app_fit_nl_bias}), where the impact is larger due to the smaller angular scale covered by the simulation window.

In the baseline application, we consider fits only to scales $\theta > 10'$ for both the clustering and galaxy-galaxy lensing, and $\theta > 5', 20'$ for the cosmic shear $\xi_+$ and $\xi_-$ respectively. These cuts serve two purposes. First, they remove scales in which the non-linear bias fit used for the noise-free theory curve as described in Appendix \ref{sec:app_fit_nl_bias} show the greatest divergences between the simulations and the theory. Secondly, and most importantly, they define realistic scales on which the data vector is fit in real data. These choices correspond roughly to the scale cuts used in the analysis of the DES Year 3 data vector \citep{DESY3ClGGL, DESY3Shear}. However, we note that in practice the scale cuts applied typically are redshift dependent, and are defined by intensive investigations on the sensitivity of the data to systematics (such as non-linear galaxy bias and baryonic effects) at a given angular scale. 
Such a detailed investigation on required cuts for the analysis of generation Stage-IV surveys is considered beyond the scope of this work, but should be considered in detailed survey-specific investigations. To investigate the importance of scale cuts on the sensitivity to magnification, we consider an additional optimistic case where these cuts are all halved, and thus bring in extra small scale data into the data vector (labelled "scale case 2").

For the cases where all magnification (all mag) and no magnification (no mag) terms are included, contours are constructed from nested sampling chains using {\sc MultiNest} \citep{Multinest1}, where convergence is defined relative to changes in the evidence $\mathcal{Z}$ as the point where $\Delta \log \mathcal{Z} < 0.5$. Contour corner plots are constructed using {\sc ChainConsumer} \citep{ChainConsumer}.

For the cases where only position and shear magnification are included, we do not produce contours but instead utilise {\sc SciPy} annealing \citep{SciPy} with Powell optimisation to determine the maximum-a-posteriori (MAP) point.

We note that all inference methods may suffer from an ``optimisation bias'' which results from limitations in the method in which the posterior is probed. For MCMC applications, we term this as a ``sampling bias'', where the noise from the finite sampling of the posterior can induce shifts in the inferred posterior and its maximum. Whilst this sampling bias also affects the `all mag' case and differs in size and direction to that in the `no mag' case, we consider it to be small and therefore neglect it in the interpretation (see Section \ref{sec:results} for further discussion).

We fit for three cosmological parameters, the dark energy equation of state ($w$), the cold dark matter density ($\Omega_{c}$) and the linear dark matter clustering amplitude ($\sigma_8$). These three parameters cover the primary parameters of interest in current and future weak lensing surveys but we note that extended parameter sets will be fit in future data, and that the shape of posteriors and size of cosmological bias will be closely linked to the parameter set and the freedom of the model to vary. We consider this reduced set for simplicity in visualisation and in compute complexity. Fiducial values for each are given in Table \ref{tab:priors}. As we do not allow for variation in the baryon density $\Omega_{\rm b}$, the results for  $\Omega_{\rm c}$ may be interpreted as those on the total matter density, $\Omega_{\rm m} = \Omega_{\rm c}+\Omega_{\rm b}$, where the contours and MAP values are shifted by the fiducial value for $\Omega_{\rm b}$ (note that biases are unchanged between both parameterisations).

In clustering analyses, the magnification effect can be considered as a spatially-dependent alteration to the redshift distribution for each tomographic bin \citep[see][for an application of measurements of mean redshift to determine cluster profiles]{CouponRedshiftShift}. It is therefore reasonable to expect that where an analysis allows for freedom in altering the redshift distribution,  these will be sensitive to the magnification and therefore could impact cosmological sensitivity through covariance between these parameters. We therefore allow for the simultaneous inference of $10$ parameters determining shifts in the mean of the redshift distributions. Alongside these, we also allow for freedom in the linear galaxy bias, which we parameterise as a simple linear bias with $b = b_0 + b_zz$. We do not allow for variation of non-linear bias parameters in this step, however these are fit as part of the first precursor step (Appendix \ref{sec:app_fit_nl_bias}), and they already contain redshift dependent terms which are fit to the SLICS-derived correlations. The linear bias model here has as its truth value $b=1$ for all redshifts (i.e. $b_0 = 1$ and $b_z = 0$), and does not tell us about the true redshift dependence of galaxy bias in the SLICS simulations. Instead, this model allows for the realistic interpretation of the impact of magnification bias in an analysis where galaxy bias is inferred, as well as allowing for the determination of bias in these parameters as a result of the magnification. Moreover, since galaxy bias will be degenerate with cosmological model parameters of interest, allowing freedom for the galaxy bias to vary increases the realism in both the inferred cosmological accuracy and precision.

The priors applied to the fit parameters are detailed in Table \ref{tab:priors}. In all cases, wide flat priors are applied to the cosmological parameters. The parameters labelling the shift in the mean of the redshift distributions in each tomographic redshift bin are given a normal prior where these are not fixed and inferred from the mock data, with mean of zero and width of $\sigma = 0.01$, similar to those applied on the Y1 analysis of the DES 3x2pt using {\sc METACALIBRATION} \citep{DESY13x2pt}.

In presenting results, we demonstrate the significance of any bias using both values for the change in $\Delta \chi^2$ and probability to exceed (p-value) for the biased point relative to the case with no magnification in the data vector in the $w_0-\sigma_8$ plane. In both cases, for simplicity we assume Gaussianity in the posteriors, so that posterior shapes can be fully described by the covariance from the MCMC chains. To verify that this assumption is reasonable, we validated these values by using affine-invariant MCMC sampling \citep[{\sc EMCEE},][]{EMCEE} to resample KDE smoothed versions of the posterior chains. Each chain was run to $10^6$ links (convergence was not explicitly enforced), and confidence levels derived from these MCMC chains without assuming Gaussianity. It was found that the p-values produced in this way agreed very well with those using a Gaussianity assumption, and made no qualitative difference to conclusions. To limit the impact of potential sampling bias, the $\Delta \chi^2$ and p-value were determined using the true input cosmology parameters, not those inferred from the no-magnification chains.

\begin{table}
\begin{center}
\begin{tabular}{|c|c|c|}
\hline
\hline
Parameter & Prior & Fiducial Value \\
\hline
$\sigma_8$ &  $U(0, 1.2)$ & 0.826\\
$w_0$ & $U(-1.5, 0)$ & -1\\ 
$\Omega_c$ & $U(0, 1.5)$ & 0.243\\
$\Delta\langle z_i \rangle$& $\mathcal{N}(0,0.01)$ & 0\\
$b_0$ & $U(-3, 5)$ & 1 \\
$b_z$ & $U(-3, 5)$ & 0\\
\hline
\hline
\end{tabular}
\caption{Priors applied to and fiducial values for each parameter, including cosmological model parameters ($\sigma_8, w_0, \Omega_c$), redshift distribution calibration parameters as a shift in the mean of bin $i$ ($\Delta \langle z_i \rangle$) and galaxy bias parameters ($b_0$, $b_z$). $U(x,y)$ indicates a flat uniform prior with support between $x$ and $y$. $\mathcal{N}(\mu, \sigma)$ indicates a Normal prior with mean $\mu$ and width $\sigma$. }\label{tab:priors}
\end{center}
\end{table}

\subsection{Limitations of the analysis}\label{sec:limitations}

The emulation and inference process described above provides a clean and low cost method for the determination of the sensitivity of future 3x2pt analyses to the magnification of all aspects of the data vector. In particular, the emulation process allows for the determination of the additional correlations due to magnification without the need to utilise source-specific magnified positions and fluxes, and as such allows such terms to be determined without observational uncertainties such as spatially varying survey selection (for example due to survey depth variation, foreground obscuration and dust extinction) or photometric uncertainty \citep[e.g.][]{HildebrandtMagSystematics}. Moreover, the emulation process allows for the use of simulation suites which do not contain magnification-adjusted source position or flux. The inference process allows for a nearly noise-free determination of the model parameter bias, since the process of differencing simulation-measured correlation functions induces a level of noise cancellation in the magnification correlations, thus disentangling the parameter bias from the realisation specific shifts in the posterior.

However, the process described utilises a series of simplifications and assumptions which limit its ability to accurately predict biases for specific surveys:
\begin{enumerate}
\item{We consider only a fixed magnification strength with redshift, whilst in practice the magnification strength varies with magnitude limit and redshift due to survey selection considerations and evolution of the underlying galaxy luminosity function.}
\item{The emulation method assumes that modelling a well-known, simple, hard selection (and therefore well-known $\alpha$) on the sources is valid and does not consider the impact of more complicated selections due to e.g. photometric uncertainty, colour-colour selection, obscuration, reddening, extinction or survey depth variation \citep[see, e.g.][]{HildebrandtMagSystematics,GGL_Obscuration,Morrison15}. These can be spatially varying and thus degenerate with the magnification signal or complicate the accurate inference of the magnification strength. Further, source weighting used in shear analysis will implement an extra effective selection on the shear sample, which may vary with redshift and magnitude.} 
\item{The emulation assumes the valid application of the weak-lensing limit, and therefore the linearisation of magnification and reduced shear re-weighting.}
\item{We assume identical source selection in both the position and shear sample, where in practice these are likely to differ due to different requirements on e.g. photometric redshift determination and signal-to-noise.}
\item{Whilst we assign photometric redshift errors, these are limited to those sampled from a normal distribution and do not contain outliers.}
\item{We consider a reduced $w$CDM cosmology where only the parameters $w_0, \sigma_8, \Omega_c$ are allowed to vary, alongside a simple linear galaxy bias and shift in the mean of each tomographic redshift distribution.}
\item{We utilise redshift independent scale cuts, where in practice different scales will be considered acceptable due to redshift dependent systematic bias and the evolution of co-moving scale cuts (such as those due to non-linear clustering or baryonic effects).}
\end{enumerate}
In particular, the use of simulations which include magnified source positions and magnitudes, and realistic redshift distributions alongside other systematics effects could be used to investigate in detail not only the final sensitivity of the analysis to the magnification, but also the ability to mitigate these effects with techniques that will be used in the final analysis of the data.

\section{Results}\label{sec:results}

\begin{table*}
\begin{tabular}{l|cccc|cccc|cccc}
\hline
\hline
 & Position mag &   &  &  &  Source mag &  &  &  &  All mag &   &  &   \\\hline

Observation & $\Delta w_0$ & $\Delta \sigma_8$  & $\Delta \chi^2$  & p-value  & $\Delta w_0$  & $\Delta \sigma_8$  & $\Delta \chi^2$  & p-value  & $\Delta w_0$  & $\Delta \sigma_8$  & $\Delta \chi^2$  & p-value  \\\hline

\\
\multicolumn{13}{|c|}{\underline{Baseline, Mag + Reduced Shear}}\\
\\
3x2pt & -1.9e-02 & 1.7e-03 & 10.3 & {\bf 5.9e-03} & 1.5e-03 & 5.5e-03 & 20.2 & {\bf 4.1e-05} & -4.7e-02 & 5.5e-03 & 77.9 & {\bf 1.2e-17} \\
2x2pt (auto) & -6.9e-02 & 1.1e-02 & 4.5 & {\bf 1.1e-01} & 2.3e-04 & 1.3e-03 & 0.0 & 9.8e-01 & -6.3e-02 & 1.1e-02 & 4.3 & {\bf 1.2e-01} \\
gg & - & - & - & - & - & - & - & - & 1.4e-02 & 7.6e-03 & 16.0 & {\bf 3.3e-04} \\

\\
\multicolumn{13}{|c|}{\underline{Baseline, Mag Only}}\\
\\
3x2pt & -4.1e-02 & 1.5e-05 & 18.8 & {\bf 8.1e-05} & -1.3e-03 & 1.8e-03 & 2.6 & {\bf 2.7e-01} & -5.9e-02 & 1.4e-03 & 51.2 & {\bf 7.7e-12 } \\
3x2pt (auto) & - & - & - & - & - & - & - & - & -8.2e-02 & 2.7e-03 & 61.6 & {\bf 4.1e-14} \\
2x2pt (auto) & -6.6e-02 & -2.2e-03 & 1.2 & {\bf 5.4e-01} & -8.4e-04 & 5.3e-04 & 0.0 & 1.0e+00 & -9.2e-02 & -1.3e-03 & 2.4 & {\bf 3.0e-01} \\
gg  & - & - & - & - & - & - & - & - & 5.1e-03 & 2.8e-03 & 2.1 & {\bf 3.5e-01} \\
ng & -3.0e-01 & 7.8e-02 & 11.8 & {\bf 2.7e-03} & 7.2e-03 & 3.3e-04 & 0.0 & 1.0e+00 & -2.6e-01 & 8.0e-02 & 10.3 & {\bf 5.9e-03} \\
nn (auto) & - & - & - & - & - & - & - & - & -1.3e-01 & -2.2e-01 & 6.8 & {\bf 3.4e-02} \\

\\
\multicolumn{13}{|c|}{\underline{Fixed z, Mag Only}}\\
\\
3x2pt & - & - & - & - & - & - & - & - & 5.7e-03 & 3.3e-03 & 3.3 & {\bf 1.9e-01} \\
2x2pt (auto) & - & - & - & - & - & - & - & - & -6.7e-02 & 1.4e-02 & 5.1 & {\bf 7.7e-02}  \\
\\
\multicolumn{13}{|c|}{\underline{Scale case 2,  Mag Only}}\\
\\
3x2pt& -2.0e-02 & 2.6e-03 & 20.9 & {\bf 2.8e-05} & -8.9e-04 & 1.8e-03 & 2.9 & {\bf 2.4e-01} & -3.6e-02 & 2.4e-03 & 39.2 & {\bf 3.0e-09} \\
gg & - & - & - & - & - & - & - & - & 6.8e-03 & 2.9e-03 & 2.6 & {\bf 2.7e-01}  \\

\\
\multicolumn{13}{|c|}{\underline{L21 $\alpha$, Mag + Reduced Shear}}\\
\\
3x2pt & 6.2e-02 & 1.1e-02 & 43.5 & {\bf 3.6e-10} & -3.2e-02 & 1.0e-02 & 127.5 & {\bf 2.0e-28} & 2.0e-02 & 2.1e-02 & 254.6 & {\bf 0.0e+00} \\
2x2pt (auto) & 1.7e-03 & 3.1e-02 & 26.9 & {\bf 1.5e-06} & 4.4e-03 & 4.3e-03 & 0.5 & 7.8e-01 & 4.3e-03 & 3.5e-02 & 34.2 & {\bf 3.7e-08} \\
gg & - & - & - & - & - & - & - & - & 4.4e-02 & 2.1e-02 & 118.1 & \bf{2.2e-26} \\
\hline
\hline
\end{tabular}
\caption{Parameter biases for $w_0$ and $\sigma_8$ as a function of experiment, and included magnification correlations. Also included are the change in chi-squared ($\Delta \chi^2$) and probability to exceed ($p$) for the biased parameter values with respect to the contours which do not contain a magnification contribution. The probability to exceed is defined as the probability that a random sample chosen from the un-magnified posterior lies outside the iso-posterior region on which the biased parameter values reside. Larger $\Delta \chi^2$ and smaller $p$ represent more significant biases.  Values in bold represent cases where the p-value indicates $>1\sigma$ biases in the $\sigma_8-w_0$ plane. Cosmic shear is labelled gg, galaxy-galaxy lensing ng, and clustering nn. 2x2pt labels the combination of nn+ng, 3x2pt the further inclusion of gg. The ``auto'' label indicates that the clustering (nn) data vector elements contained only auto-correlations in redshift bins. Note that $p=1$ and $p = 0$ may be obtained from very small or large $\Delta \chi^2$ values, and result from numerical limitations in the calculation of the p-value.}\label{tab:biases_and_tension}
\end{table*}

\begin{figure*}
\includegraphics[width=\textwidth]{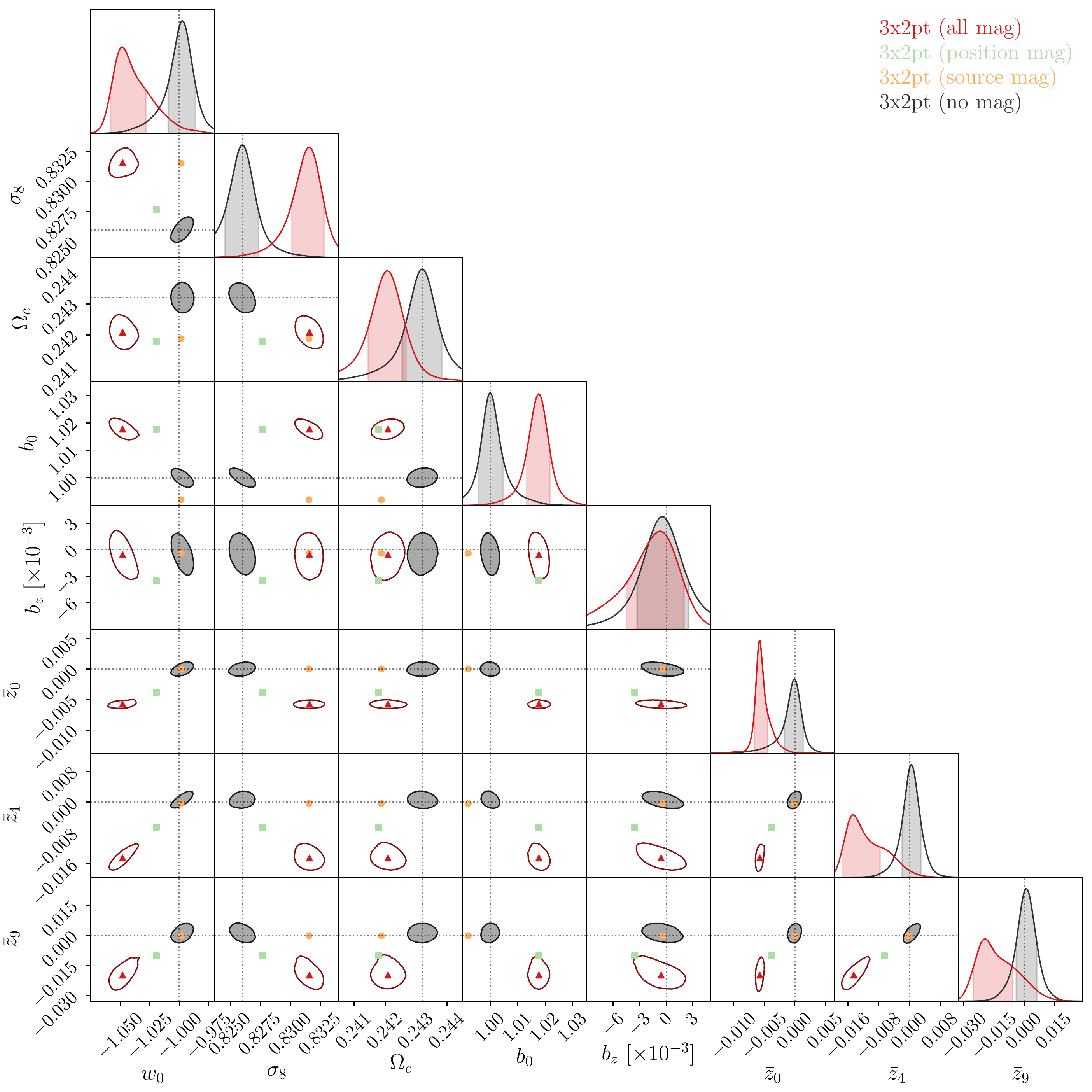}
\caption{Inferred parameter contours for the case where the full 3x2pt data vector is analysed. This corresponds to the results for the 3x2pt analysis under the `Baseline' header in Table \ref{tab:biases_and_tension}. All contours are shown at $1\sigma$ level. Grey contours correspond to a fit where no magnification is included in the data vector. Red contours include all magnification contributions in the data vector, and the red triangle shows the corresponding MAP point. The green square shows the max-posterior point when only the position sample is magnified, and the orange circle when only the shear sample is magnified. Note that shifts in the mean redshift of each tomographic bin is allowed, although only 3 are shown here for clarity.}\label{fig:contours_3x2pt}
\end{figure*}

\begin{figure*}
\includegraphics[width=\textwidth]{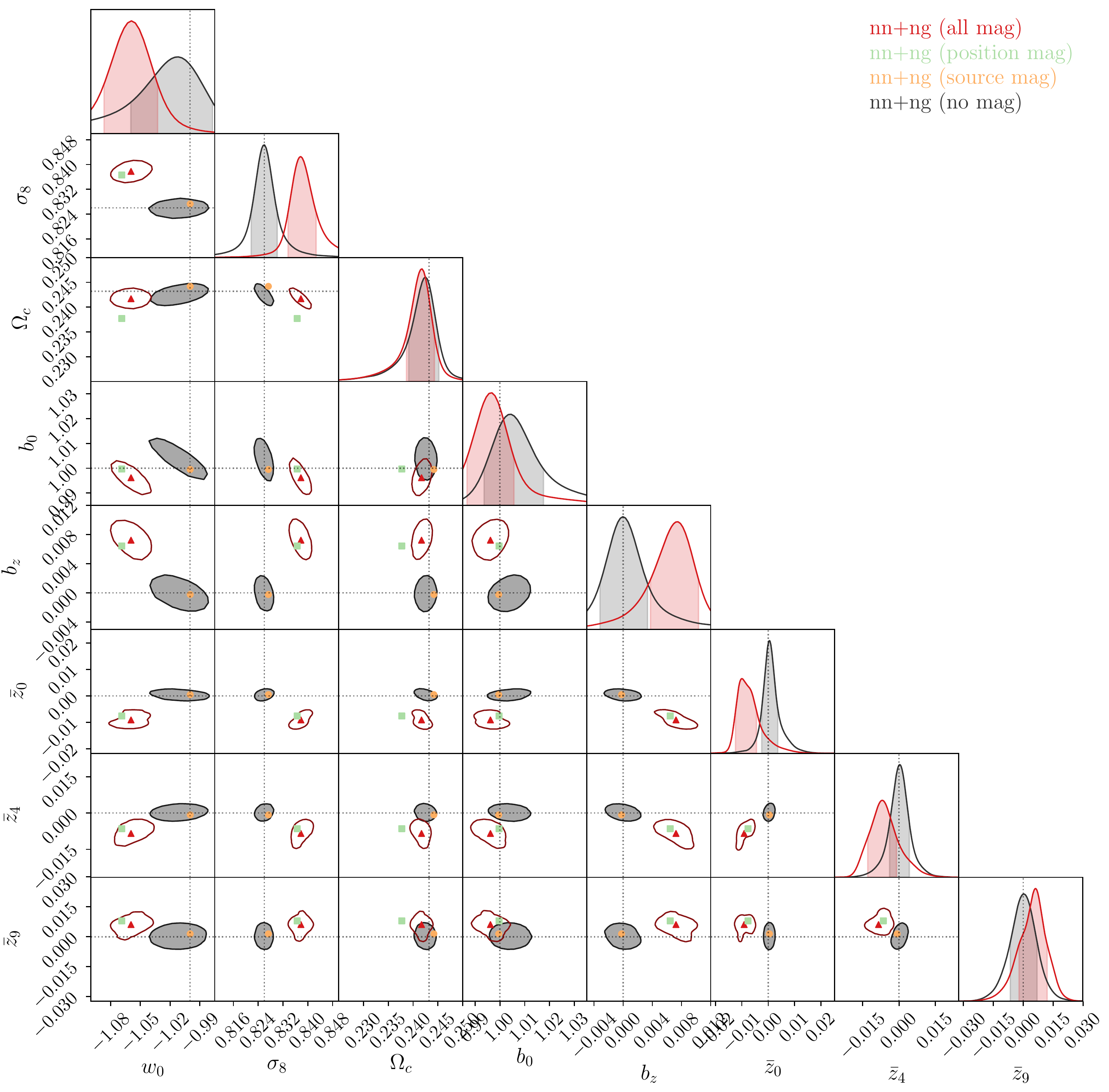}
\caption{As Figure \ref{fig:contours_3x2pt}, for the 2x2pt analysis (clustering + galaxy-galaxy lensing), corresponding to 2x2pt under the `Baseline' heading in Table \ref{tab:biases_and_tension}.}\label{fig:contours_2x2pt}
\end{figure*}

\begin{figure*}
\includegraphics[width=0.75\textwidth]{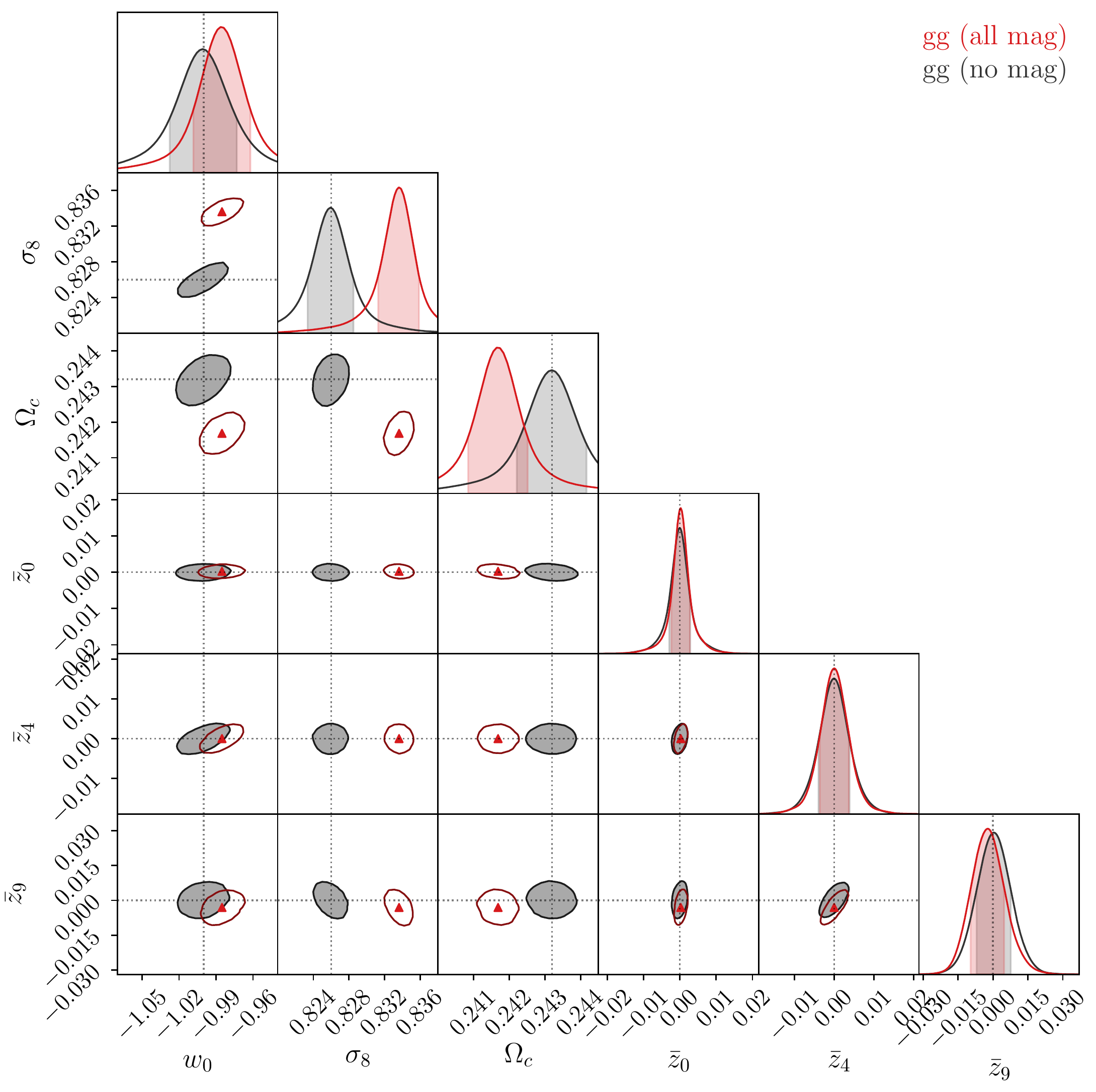}
\caption{Inferred parameter contours for the case where the cosmic shear only data vector is analysed. This corresponds to the results for the gg analysis under the `Baseline' header in Table \ref{tab:biases_and_tension}. All contours are shown at $1\sigma$ level. Grey contours correspond to a fit where no magnification is included in the data vector. Red contours include all magnification contributions in the data vector, and the red triangle shows the corresponding MAP point. Galaxy bias is not included as a free parameter. Note that shifts in the mean redshift of each tomographic bin is allowed, although only 3 are shown here for clarity.}\label{fig:contours_gg}
\end{figure*}

Table \ref{tab:biases_and_tension} gives information on the biases and significance of these biases on the $w_0$-$\sigma_8$ plane. To specify significance of the bias, we present both the $\Delta\chi^2$ for the biased ML point compared to the posterior as determined from the chain from the mock data that did not include the magnification terms, alongside a probability-to-exceed ($p$) value. Note that the biases and significances are taken with respect to the truth values, not the mean or MAP of the no-magnification simulations, to minimise the impact of sampling bias in the ``no mag'' chain. Larger $\Delta\chi^2$ and smaller $p$ values correspond to more significant biases. Although these values can be determined on the full N-dimensional parameter set, we instead restrict them to the $w_0$-$\sigma_8$ plane to aid interpretation. In Table \ref{tab:biases_and_tension}, we have highlighted those p-values which indicate a $\geq1\sigma$ shift in parameter values due to magnification, however shifts below $1\sigma$ would also be a concern in application to data.

The baseline considers the full analysis presented in the preceding sections, including both the effects of magnification and reduced shear. To aid interpretation of the magnification effect specifically, we consider a variant of the baseline where the reduced shear effect has not been included under the header ``Mag Only''. Further variants include the case where the redshift distribution calibration parameters are not fit to the data, but fixed to their fiducial values (header `` Fixed z''), where more optimistic cuts are applied which reduce the minimum angular scale included in the fit (header ``Scale case 2'', see Section \ref{sec:inference}), and where the redshift dependent $\alpha$ values of \citet{Lepori} are used (header ``L21 $\alpha$''). 

Figures \ref{fig:contours_3x2pt}, \ref{fig:contours_2x2pt}, \ref{fig:contours_gg} show parametric contours for the baseline 3x2pt, 2x2pt (auto) and gg analyses (as summarised in Table \ref{tab:biases_and_tension}). These figures show corner plots for all parameters, including nuisance parameters such as galaxy bias and the mean of each redshift distribution (although each of the 10 tomographic bins is given a nuisance parameter for shifts in the mean of the redshift distribution, only 3 are plotted for clarity). We emphasise that although we show contours for the `all mag' case, that the inference process described in Section \ref{sec:inference} is approximately noise-free, and thus in deducing the significance of the bias we should only consider the position of the maximum-posterior point where magnification is included to the width of the posterior in the no-magnification case. We should not account for the width of both the ``all-mag'' and ``no-mag'' contours in interpreting tension. Vertical and horizontal bars represent the input truth value for each parameter. As discussed in Section \ref{sec:inference}, we see evidence for small shifts between the maximum-posterior values in the no-magnification case and the input truth, due to sampling bias. These are smaller than the typical shifts in inferred parameters due to magnification and therefore do not alter their interpretation.

In the following, we factor in the reduced shear as an extra contribution in the source magnification. However, this effect would be present in the data even without source magnification, and therefore would also be present in the ``position mag'' result even though it is not included here. As a consequence, when interpreting the relative importance of the magnification of the position and source sample, it is recommended to compare these in the case where reduced shear is not included.
It should also be noted that as the magnification strength $\alpha > 1$ for all redshift bins in the baseline analysis, the impact of the reduced shear is to increase the overall lensing prefactor in the shear sample (Eqn \ref{eqn:reduced_shear_weight}), and thus is expected to increase parameter biases over the magnification-only case. This is not a general case, and for survey selections where $\alpha<1$ the reduced shear would counteract the magnification and reduce parameter biases. This is the case for L21 $\alpha$ values for $z\lesssim 1$.

\subsubsection*{Cosmological biases due to un-modelled magnification}

We see that in all cases the presence of the un-modelled full magnification contribution in the data set leads to a bias which is statistically significant.

For the full 3x2pt analysis, the biases induced on both $w_0$ and $\sigma_8$ are large in comparison to the typical precision with which these parameters are inferred from the data, indicating that these parameters are biased in a highly statistically significant way. Moreover, we see that whilst the magnification of the position sample is typically dominant in the size of the induced parameter biases\footnote{Comparing position and shear magnification effects in the ``Mag only'' case}, the magnification of the source sample alone induces parameter biases which are greater than $1\sigma$, with the significance further increasing when the reduced shear effect is included.

The removal of the cross-correlations in the clustering part of the data vector causes an increase in the bias of both parameters, as well as an overall increase in the significance of the bias. This is in broad agreement with the results of \cite{ThieleDuncanAlonso} which indicated that these terms induced parameter shifts which were typically in the opposite direction to those typically induced by the remainder of the data vector. However, it should be noted that such behaviour is likely to be highly dependent on the survey characteristics and sample selection, and in particular is expected to be sensitive to photometric redshift uncertainty and outlier fraction.  
These results indicate that it cannot be assumed sufficient to simply model the additional 2-point correlations which originate from the magnification of the position sample, but that the more computationally complex 3-point contributions originating from the magnification of the shear sample must also be modelled.

Similarly, for the shear-shear correlations (gg), we also see that the magnification induces parameter biases greater than $1\sigma$, increasing when the reduced shear effect is included (as in the 3x2pt analysis). This is in general agreement with the results of \cite{Deshpande2020Cosmo} who also found a $\sim 1\sigma$ bias for dark-energy parameters (though note that use a different prescription for the magnification strength and a different choice of fiducial cosmological and nuisance parameters), and confirms the importance of mitigating or modelling these additional 3-point terms.

The 2x2pt analyses is sensitive to the magnification of the position sample, which in contrast to source magnification affects both parts of the data vector. Thus, for the 2x2pt we find that whilst magnification of the position sample must be modelled, magnification of the shear sample may be more safely ignored. This suggests that the 2x2pt analysis may be used to verify the full 3x2pt analysis, by indicating that the magnification of the position sample has been safely mitigated.

For the remaining experiments we see that typically the  magnification of the shear sample induces insignificant parameter biases. For the galaxy-galaxy lensing (ng) analysis, we see that source magnification induces a shift in $w_0$ larger than those seen in the 3x2pt analysis, however the significance of that shift is mitigated by the decreased constraining power of the base analysis and a reduction in the $\sigma_8$ bias. When further combined with clustering (nn+ng or 2x2pt), the bias in $w_0$ is reduced by an order of magnitude whilst the bias in $\sigma_8$ increases only slightly, as the clustering (which is insensitive to magnification in the shear sample) mitigates the parameter shifts due to source magnification from the galaxy-galaxy lensing part of the data vector. 

Finally, we see that the inclusion of smaller angular scales (under header ``Scale case 2'') makes no substantial qualitative or quantitative difference to the significance of cosmological biases.

\subsubsection*{The impact of redshift dependent magnification strength}

We see that the propagation of the \citet{Lepori} (L21) redshift dependent magnification strength, as derived from \emph{Euclid} Flagship simulations, induces larger biases in all experiments. The impact of these redshift dependent biases on the magnification signal is expected to be two-fold, compared to the baseline case where $\alpha = 1.3$ is used for all bins. Firstly, for $z\lesssim 1$, L21 has $\alpha  < 1$, which would allow for negative magnification corrections due to magnification of the position sample, and smaller positive correlations due to the shear sample (Eqn \ref{eqn:reduced_shear_weight} shows that for $\alpha<1$ the magnification and reduced shear effects partially cancel). Secondly, for high redshift, the L21 $\alpha$ values are much larger (reaching $\alpha \sim 2.75$ at $z=1.91$, which is also used for all bins at higher redshift in this analysis). The latter in particular would be expected to increase cosmological biases, both due to the increase in the magnification signal for these bins, but also since high-redshift bins have higher cosmological sensitivity to magnification bias \citep{ThieleDuncanAlonso}. We note that the shallower survey model used in \cite{Lepori} extends to a maximum redshift smaller than that considered here. The redshift dependence of $\alpha$ may be weaker in deeper surveys, where the apparent magnitude limit samples a flatter part of the galaxy luminosity function at high redshift. 

\subsubsection*{Biases on nuisance parameters and impact on cosmology}

The presence of magnification in the data vector also induces significant shift in the nuisance parameters for galaxy bias and shifts in the mean of the redshift distributions for the 3x2pt and 2x2pt analyses, but typically has little impact on redshift calibration parameters for the cosmic shear. Table \ref{tab:biases_and_tension} shows summaries in each case where the mean of the redshift distributions are fixed to their fiducial value under the header ``Mag Only, Fixed z'', and in comparison to the baseline case we see that there is no consistent change in the size of cosmological parameter bias in each case. For the 3x2pt analysis we see that fixing the mean redshift of each bin reduces the bias in $w_0$ by an order of magnitude, whilst for the 2x2pt doing so increases the bias in $\sigma_8$ by an order of magnitude. This indicates that the sensitivity of the analysis to magnification is itself reliant on the form of the redshift distribution and the freedom in parameterising it when inferred from the data, and thus care must be taken in application to data. This is likely more important where outlier redshifts are present, which can complicate interpretation of the magnification to the intrinsic signal, and where the redshift distributions are given freedom to vary in shape as part of the inference, e.g. when fitting for cosmology or for the redshift distributions themselves using clustering. Whilst we consider galaxy bias as a nuisance parameter in this paper, the induced bias in the inferred galaxy bias indicates that magnification must also be considered in analyses which aim to infer and interpret the galaxy bias.

\subsubsection*{Summary}
These results emphasise that magnification must be modelled to avoid significant parameter bias in future weak lensing surveys (also confirming and reinforcing the conclusions of the literature referenced in Section \ref{sec:intro}). They also indicate that it is not adequate to only model the effect of magnification of the position sample, and that magnification of the shear sample induces significant parameter biases in both 3x2pt and cosmic shear analyses. The exception to this is in the 2x2pt, where the magnification of the shear sample is a subdominant effect to the highly significant position sample magnification. The complicated interdependence of the cosmological bias on the parameter set chosen, particularly redshift distribution calibration parameters, and the impact of survey specific galaxy selection on the accuracy of the derived magnification strength, emphasises that sensitivity to and mitigation of magnification must be tested in survey-specific analyses.

\section*{Conclusions}

In this paper, we have utilised the high-resolution SLICS simulation suite to emulate and analyse the cosmological impact of the magnification of both the position and shear samples for a modern Stage-IV-like weak lensing survey. The emulation of the magnification effects was achieved at the correlation-measurement stage by re-weighting each source with a weight mimicking the effective change in number density of each source in a flux-limited survey in the presence of a magnification field. The additional correlations due to magnification were isolated by subtracting the same measurement without this re-weighting applied. The process of emulating the magnification effect allows for the determination of the additional correlations in each aspect of the weak lensing data vector (including cosmic shear, galaxy-galaxy lensing and clustering) resulting from the magnification of the galaxy samples and the reduced shear, without the presence of magnified galaxy flux and positions in the simulation catalogues. This therefore has the advantage that these effects may be emulated at low cost, without theoretical modelling uncertainties, and without observational uncertainties such as variable survey selection effects. The impact of the magnification and reduced shear could therefore be determined cleanly.

We analysed the impact of these terms in inferred cosmological and nuisance parameters, when the fit theory term does not include a theoretical modelling for each. We have shown that the impact of magnification and reduced shear induces large, statistically significant biases in cosmological parameters in all aspects of the data vector, indicating that it is imperative that the magnification is modelled in all forms of analysis with Stage-IV surveys. For example, the dark energy equation of state shows parameter biases of $~1-10\%$ across cosmic shear, 2x2pt, 3x2pt and clustering analyses, raising to $~25\%$ in the galaxy-galaxy lensing alone. Such biases could lead to a measurement showing spurious tension with $\Lambda$CDM cosmology. This confirms the conclusions of e.g. \citet[][]{ThieleDuncanAlonso, Lepori, Deshpande2020Cosmo} and others mentioned in the introduction, and shows for the first time that this is true in the combined 2x2pt (galaxy-galaxy lensing and clustering) and 3x2pt (plus cosmic shear) analyses when reduced shear and magnification of the source sample is included.

By turning off the emulation of the magnification of the position sample as a proxy for perfect mitigation through modelling, we show that the magnification of the shear sample alone induces significant biases ($>1\sigma$ in the $w_0-\sigma_8$ plane) for both the cosmic shear and 3x2pt analyses, but not for the 2x2pt. Both the magnification of the shear sample, and the reduced shear, induce 3-point corrections to the shear and galaxy-galaxy lensing 2-point correlations which may be computationally complex or expensive to compute \citep[see e.g.][for further discussion]{Deshpande2020Cosmo}. Consequently, mitigating these effects through modelling will complicate Monte-Carlo based inference, and thus this result indicates that effort on efficiently implementing these corrections is of high importance for future analyses.

We have shown that the inference of shifts in the tomographic redshift distributions are sensitive to the magnification in the 3x2pt and 2x2pt analyses, and that the observed biases in cosmological parameters are sensitive to the freedom to alter the redshift distributions in an non-trivial way. This indicates that whilst the observed biases on cosmological parameters of interest may depend on the freedom to alter the inferred redshift distributions, large shifts in these distributions may be indicative of a poorly mitigated or modelled magnification contribution. By contrast, an analysis using only the cosmic shear signal does not show significant sensitivity in the redshift distributions.

It has been widely demonstrated that magnification of the position sample is a major systematic for clustering analyses, and those which combine clustering with other weak lensing measurements. Recent investigations have shown that modern lensing surveys are entering the period where their statistical power requires that magnification of the shear sample in a cosmic shear analysis must also be considered \citep{Deshpande2020Cosmo}, alongside magnification of the position sample \cite[e.g.][]{ThieleDuncanAlonso}. This paper shows that in implementation the magnification can no longer be considered an effect which impacts mainly clustering analyses, but instead must be considered a general lensing effect which enters all aspects and combinations of the data vector in modern weak lensing analyses. As such, it is imperative that survey-specific investigations are utilised to demonstrate and test specific modelling and mitigation strategies for magnification alongside other commonly considered systematics.

\section*{Acknowledgements}

This paper extends work undertaken as a Master's project by Alexander Langedijk, on emulating the additional correlations due to magnification in the galaxy-galaxy lensing observable with SLICS.
CAJD acknowledges support from the Beecroft Trust. JHD acknowledges support from an STFC Ernest Rutherford Fellowship (project reference ST/S004858/1). Post-processing computations of the SLICS simulations were carried on the Cuillin cluster at the Royal Observatory of Edinburgh, which is managed by Eric Tittley.

{\footnotesize Every author has contributed to the development of this paper. CAJD has led the analysis, JHD has developed the numerical simulations, LM developed the concept of the simulation-based emulation.}




\bibliographystyle{mnras}
\bibliography{references}




\appendix


\section{Fitting non-linear bias to the simulations}\label{sec:app_fit_nl_bias}

As detailed in Section \ref{sec:inference}, we infer the parameter precision and bias in a quasi-noise-free process where the simulation-measured magnification terms are added to a theory-derived data vector of correlations without the magnification included. Since both the intrinsic correlations and the simulation measured magnification terms are dependent on the galaxy bias, we require that the form for the redshift and scale dependent bias implemented in the theory derived no-magnification data vector matches that of the simulations. As a first step, we therefore directly infer a form for the galaxy bias parameterised as 
\begin{equation}\label{eqn:nl_bias}
    b(k,\langle z \rangle) = (A_2k^2+A_1k+A_0)/(1+Bk)
\end{equation}
 with all coefficients linear relations on the mean redshift of each bin, by fitting to the simulations themselves for both the clustering and galaxy-galaxy lensing terms. The simulations do not contain a contribution from the magnification. The clustering is limited to the auto-correlations, which correspond to the highest signal-to-noise measurements in the full clustering data vector, and the galaxy-galaxy lensing uses all correlations where the shear sample is in a higher redshift bin than the lens sample. Using both the clustering and ggl data allows for the optimal inference of the galaxy bias terms, by including terms which are both linear in the galaxy bias (from the galaxy-galaxy lensing) and quadratic (from the clustering auto-correlations). 

\begin{figure*}
\includegraphics[trim = 100 50 20 80, clip, width=\textwidth]{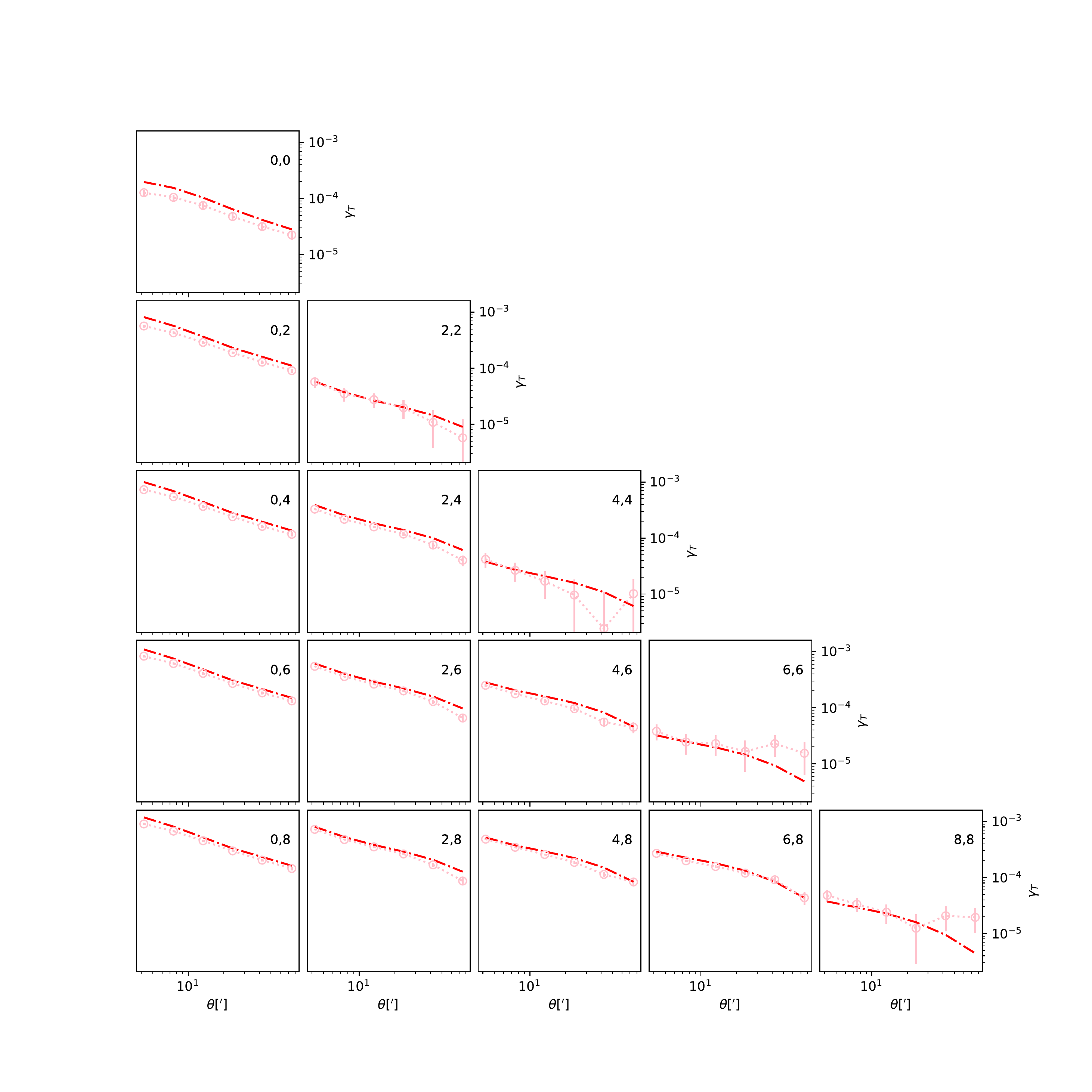}
\caption{The galaxy-galaxy lensing signal as measured across 10 lines-of-sight in the SLICS LSST simulations (pink, circles) and the theory curves corresponding to the best fit non-linear bias model (red, dashed). Error bars shown are auto-variance terms in the analytic covariance. Note that circle markers are chosen to aid visualisation, and may be larger than the error bars. Only every second redshift bin is chosen for visualisation purposes. Redshift bin for the position sample increases top-to-bottom, and edshift bin for the shear sample increases left-to-right.}\label{fig:app_ng_sims_bias_fit}
\end{figure*}

\begin{figure*}
\includegraphics[width=0.85\textwidth]{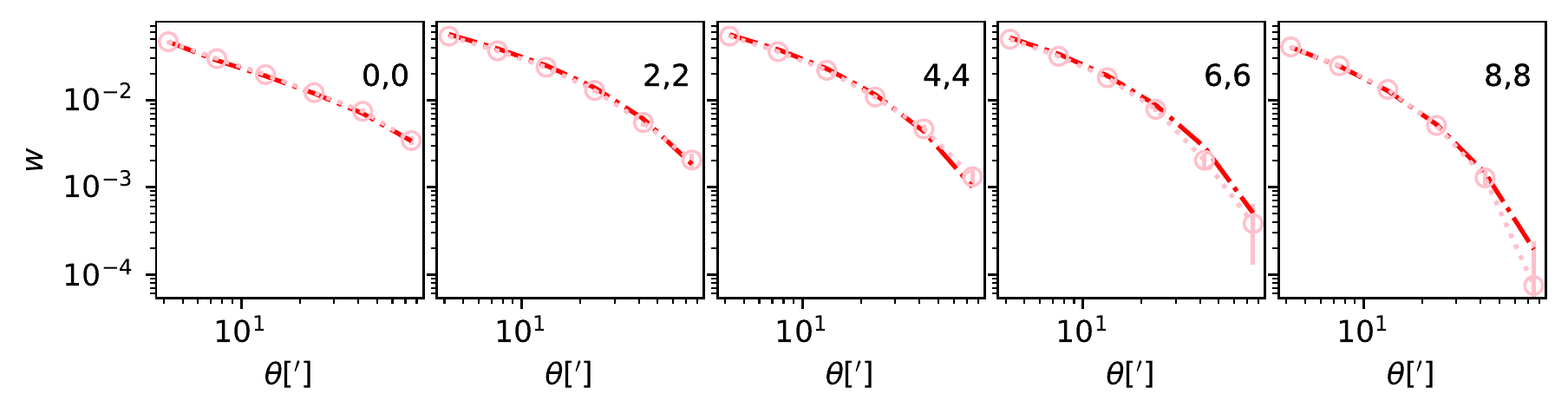}
\caption{The clustering signal as measured across 10 lines-of-sight in the SLICS LSST simulations (pink, circles) and the theory curves corresponding to the best fit non-linear bias model (red, dashed). Error bars shown are auto-variance terms in the analytic covariance. Note that circle markers are chosen to aid visualisation, and may be larger than the error bars. Only every second redshift bin is chosen for visualisation purposes.}\label{fig:app_nn_sims_bias_fit}
\end{figure*}

We follow the inference process as detailed in Section \ref{sec:inference}, where cosmology and redshift distribution parameters are fixed to their fiducial values, and only bias terms detailed in eqn \ref{eqn:nl_bias} are allowed to freely vary. The covariance is taken to be the analytically derived covariance as detailed in Section \ref{sec:inference}, containing Gaussian, non-Gaussian and super-sample covariance (SSC) contributions. As the data vector is averaged over 10 lines-of-sight, each covering 100 sq-deg, the Gaussian and non-Gaussian terms of the covariance are scaled to 1000 sq-deg, whilst the SSC term is evaluated at the 100 sq-deg of each cone. Correlations are considered down to 3 arc-minute separations, below which resolution effects in the simulations are expected to become dominant. The resulting maximum-likelihood fit is demonstrated in Figs \ref{fig:app_ng_sims_bias_fit} and \ref{fig:app_nn_sims_bias_fit}, for the galaxy-galaxy lensing and clustering respectively. Plotted error bars are given by the auto-covariance, and thus ignore covariance between data points, however the fit is taken utilising the full covariance.

\begin{table}
\centering
\begin{tabular}{ |c|r| } 
 \hline
 \hline
 $A_2$ & $-0.15\langle z \rangle - 0.46 $ \\
 $A_1$ & $-0.30\langle z \rangle +  4.92$ \\
 $A_0$ & $0.62\langle z \rangle + 0.92$ \\
 $B$ & $-1.88\langle z \rangle + 5.00$ \\
 \hline
 \hline
\end{tabular}
\caption{Best fit parameters for non-linear bias as given by Eqn \ref{eqn:nl_bias}}
\label{table:nl_bias_best}
\end{table}

The best fit parameters are given in Table \ref{table:nl_bias_best}, giving $\chi^2_{\rm red} = 1.08$ on $390$ simulation data points. Note that only every second redshift bin is plotted to aid visualisation, however all $10$ redshift bins are used as part of the fit. We see that the fit theory is in good agreement with the simulations across the whole data vector, with the largest deviations seen in the galaxy-galaxy lensing terms with the position sample at low redshifts. We emphasise however that the fit theory derived from this step is only used to set the fiducial noise free data vector in the subsequent steps. Since in determining cosmological biases this fit theory is assumed for the magnification-free data vector (on top of which the magnification terms are added) as well as the fit non-magnification theory, inaccuracies in the fit bias are expected to be of second order in the inferred cosmological biases. Therefore, whilst higher-order redshift polynomials may further improve this fit, the fiducial theory does not need to be as accurate as the more complete case where cosmological parameters, magnification strength and non-linear bias are inferred simultaneously from the simulations, and this fit is considered sufficient.

\bsp	
\label{lastpage}
\end{document}